\documentclass[showpacs,preprintnumbers,amsmath,amssymb, nofootinbib]{revtex4}
\usepackage{graphicx}% Include figure files

\usepackage{bm}% bold math

\newcommand{\be}{\begin{equation}}
\newcommand{\en}{\end{equation}}
\newcommand{\bea}{\begin{eqnarray}}
\newcommand{\ena}{\end{eqnarray}}
\begin{document}
\title{Effective gravitational equations on brane world with induced gravity described by $f(R)$ term}
\author{Joel Saavedra and Yerko V\'asquez}
\address{Instituto de F\'{\i}sica, Pontificia Universidad Cat\'olica de
Valpara\'{\i}so, Casilla 4950, Valpara\'{\i}so, Chile.}

\date{\today}

\begin{abstract}

 In this article we study a generalization of Dvali-Gabadadze-Porrati scenarios, where
 the effective theory of gravity induced on the brane is given by
 a $f(R)$ term. We obtain the
 effective gravitational equations and the effective FLRW cosmological
equation on the brane of this model. We show that this
generalization has also two regime, a $5D$ regime a low energies
that has a self-accelerated branch of interest for cosmology and a
$4D$ regime at high energies that it is described a modified
gravitational theory.

\end{abstract}


\maketitle

\section{\label{sec:level1} Introduction}

The idea of extra-dimensional theories have a long history that
begins with an original idea of Kaluza and Klein \cite{kaluza} and
finds a new realization within the modern string theory
\cite{Font:2005td, Chan:2000ms}. In particular, the
Randall-Sundrum scenario has acquired a great attention in the
last decade \cite{Randall:1999ee, Randall:1999vf}. From the
cosmological point of view, brane worlds offer a novel approach to
our understanding of the evolution of the universe, for instance
see Ref.\cite{rm} and the references therein. On the other hand,
Dvali-Gabadadze-Porrati (DGP) brane world model
\cite{Dvali:2000rv}, whose gravity behaves as four-dimensional at
short distance scale but it shows higher-dimensional nature at
larger distances, has attracted a great interest in the last time
\cite{del Campo:2007zj, Gregory:2007xy,Lue:2005ya}. This brane
world model is characterized by the brane on which the fields of
the standard model are confined, and contains the induced
Einstein-Hilbert term. It also exhibits several cosmological
features \cite{Dvali:2001gm, Deffayet:2001uk, Deffayet:2001pu}.
For example, in Ref. \cite{Deffayet:2001pu}, the authors consider
a five-dimensional bulk and obtain an accelerated expansion of the
universe at late epoch, without need to introduce the cosmological
constant. Therefore, the DGP model can be seen as providing a new
mechanism to explain the late acceleration of the universe, based
on a modification of gravitational theory \cite{Gabadadze:2007dv}
that arises from the extrinsic curvature. This new mechanics to
explain cosmological late acceleration it is more interesting than
introduce an exotic context of matter (Dark energy problem) and it
is able to explain the origin of dark energy from the
gravitational theory. However if we accept DGP as our
gravitational theory we still have some problem in order to
explain all history of the universe and match in a smooth way
different phases in the evolution of the Universe. In this sense
modified gravity emerge as serious candidate that it is able to
addressing the definitive answers to several fundamental question
about of dark energy \cite{Nojiri:2006be}. For example, the origin
of dark energy may be explained by term could be relevant at late
times. Also can be considered as the source of early time
inflation. Therefore modified gravity it is  a natural scenario to
have a theory that unified and explain both, the inflationary
paradigm and the dark energy problem. Besides, modified gravity is
expected to be useful in high energy physics. For a complete list
of the main virtues of modified gravity see Ref.
\cite{Nojiri:2006ri} were authors reviewed various modified
gravities as gravitational alternative for dark energy.
Specifically, they considered the versions of $f(R)$, $f(G)$ or
$f(R,G)$ gravity, model with non-linear gravitational coupling or
string-inspired model with Gauss-Bonnet-dilaton coupling in the
late universe where they lead to cosmic speed-up, however all
models were considered in four dimensions. Another remarkable
point is that at the intermediate epoch the gravity may be
approximated by General Relativity. In this sense it is
interesting to consider both: the DGP model and modified gravity
as an unified scenario that we hope could be able to describe the
majority of cosmological puzzles: the late acceleration,
unification of early acceleration (inflation) with dark energy,
also unification of the dark matter and dark energy and maybe a
natural scheme that provide a mechanism in order to describes the
transition from decceleration to acceleration in the universe
evolution. However, in the DGP model, the bulk is a flat Minkowski
space-time, but a reduced gravity term appears on the brane
without tension. The origin of this reduced term, proportional to
the Ricci Scalar (case of induced gravity), is understood as
coming from from quantum effects of matter fields. In this sense,
these terms should represent only the leading pieces of an
effective action, that could also contain higher-order terms in a
derivative expansion and higher powers of curvature tensors on the
brane, such as $R^2$. Such nonlinear terms are generically
suppressed by extra powers of the AdS curvature length scale, $l$,
thus at distances much larger than $l$, we expect that these
higher-order terms in the brane action could be neglected. This
kind of modification was considered  in Refs. \cite{Nojiri:2004bx}
where creation of the inflationary brane universe in 5d bulk
Einstein and Einstein-Gauss-Bonnet gravity was considered which
one surface term where terms containing higher powers of the
curvature, and recently in Ref. \cite{Atazadeh:2007gs}, where a
brane with a scalar field and curvature corrections in DGP brane
cosmology has been studied and projected equations on the brane
were obtained using a Friedmann line element in the bulk. The plan
of the paper is as follows: In Sec. II, we derive the effective
equations on the brane, as a summary of the method described in
Refs. \cite{Shiromizu:1999wj} and \cite{Maeda:2003ar}. In Sec.
III, we derive the effective equations on the brane for modified
induced gravity. In Sec. IV, we obtain the effective FLRW
cosmological equation on the brane. Finally, we present
conclusions in Sec. IV.

\section{\label{sec:level2} Effective equations on the brane; The
 Method}

In this section we summarize the main results of the
Shiromizu-Maeda-Sasaki approach \cite{Shiromizu:1999wj}. This
derivation was generalized in Ref. \cite{Aliev:2004ds}, where was
obtained the effective field equations on and off a 3-brane, to
the case where an arbitrary energy-momentum tensor in the bulk is
considered. In this brane world scenario, the observable universe
is a $4D$ surface, the brane, $M$, endowed with the metric $g_{\mu
\nu }$ (Greek indices run over $0,1,2,3$), that is embedded in a
five-dimensional space-time, the bulk, $V$, described by the
metric $g_{AB}^{(5)}$ (capital indices run over $0, 1, 2, 3, 4$).
The Standard Model particles and fields are confined to the brane,
whereas gravity is free to propagate to the bulk. The induced
metric on the brane is given by $g_{AB}=g_{AB}^{(5)}-n_{A}n_{B}$,
where $n_{A}$ is the space-like unit vector normal to $M$. In this
approach, the brane and bulk metrics remain general. The main idea
is to project the five-dimensional Einstein equations to the brane
and obtain the effective equations on the brane.

We consider the following action, \be S=S_{bulk}+S_{brane},
\label{action1} \en where $S_{bulk}$ is the Einstein-Hilbert
action in five-dimensional space-time and the brane action is \be
S_{brane}=\int_{M}d^{4}x\sqrt{-g}\left(\frac{1}{k_{5}^{2}}\,K^{\pm
}+L_{brane}\right).\label{action2} \en Here, $k_{5}$ is the
five-dimensional gravitational constant, $K^{\pm }$ is the trace
of the extrinsic curvature where $+$ and $-$ denote different
sides of the brane and $L_{brane}$ is the effective
four-dimensional Lagrangian.

Let $y$ be a Gaussian normal coordinate orthogonal to the brane, such
that the hypersurface $y=0$ coincides with the brane and $n_{\mu
}dx^{\mu }=dy$. Thus, the five-dimensional metric in terms of the
induced metric on the brane is (locally) given by
\be
ds^{2}=g_{\mu \nu }dx^{\mu }dx^{\nu }+dy^{2}.\label{metric1}
\en
The five-dimensional Einstein equation (including explicitly the
contribution of the brane) determine the five-dimensional
curvature tensor
\be
R_{AB}^{(5)}-\frac{1}{2}\,g_{AB}^{(5)}\,R^{(5)}=k_{5}^{2}\left[
T_{AB}^{(5)}+\tau _{AB}\delta (y)\right], \label{R1} \en where
$T_{AB}^{(5)}$ is the energy-momentum tensor of bulk matter fields,
\be
T_{AB}^{(5)}=-2\,\frac{\delta ^{(5)}L_{m}}{\delta ^{(5)}g^{AB}}%
+g_{AB}^{(5)}L_{m}^{(5)},\label{t1} \en and $\tau
_{AB}=g^\mu_A\,g^\nu_B\,\tau _{\mu \nu }$ and
$g^\mu_A=\delta^\mu_A$, is the effective energy-momentum tensor
localized on the brane, \be \tau _{\mu \nu }=-2\,\frac{\delta
L_{brane}}{\delta g^{\mu \nu }}+g_{\mu \nu }L_{brane}. \label{q1}
\en The delta function in (\ref{R1}) ensures that the Standard
Model fields are confined to the brane.

The Gauss equation gives the four-dimensional curvature tensor in
terms of the projection of the five-dimensional curvature, with
extrinsic curvature corrections, \be ^{(4)} R_{\beta \gamma \delta
}^{\alpha }=\left.^{(5)}R_{BCD}^{A}\right.g_{A}^{\alpha }g_{\beta
}^{B}g_{\gamma }^{C}g_{\delta }^{D}+K_{\gamma }^{\alpha }K_{\beta
\delta }-K_{\delta }^{\alpha }K_{\beta \gamma }\,,\label{k1} \en
where the extrinsic curvature of $M$ is denoted by $K_{\mu \nu
}=g_{\mu }^{A}g_{\nu }^{B}\nabla _{A}n_{B}$ and its trace is
$K=K_{\mu }^{\mu }$. The Codazzi equation determines the change of
$K_{\mu \nu }$ along the brane as \be D_{\nu }K_{\mu }^{\nu
}-D_{\mu }K=\left.^{(5)}R_{AB}\right.n^{B}g_{\mu }^{A}\,,
\label{codazzi} \en where $D_{\nu }$ is the covariant derivative
with respect to $ g_{\mu \nu }$. The five-dimensional curvature
tensor can be decomposed into the (traceless) Weyl tensor
$C_{ABCD}^{(5)}$ and the Ricci part, as \be
R_{ABCD}^{(5)}=\frac{2}{3}\,
\left(g_{A[C}^{(5)}R_{D]B}^{(5)}-g_{B[C}^{(5)}R_{D]A}^{(5)}\right)
-\frac{1}{6}\,g_{A[C}^{(5)}g_{D]B}^{(5)}R^{(5)} +C_{ABCD}^{(5)}\,.
\label{r2} \en Therefore, it follows that the induced Einstein
tensor in four dimensions $^{(4)}G_{\mu \nu}$ has the form
\begin{eqnarray}
^{(4)}G_{\mu \nu } &=&\frac{2k_{5}^{2}}{3}\,\left[ T_{RS}^{(5)}\,g_{\mu
}^{R}g_{\nu }^{S}+\left( T_{RS}^{(5)}\,n^{R}n^{S}
-\frac{1}{4}
\,T^{(5)}\right) g_{\mu \nu }\right] +KK_{\mu \nu }-K_{\mu }^{\sigma }K_{\nu
\sigma }  \nonumber \\
&&\qquad  -\frac{1}{2}\,q_{\mu \nu }(K^{2}-K^{\alpha \beta
}K_{\alpha \beta })-E_{\mu \nu }\,,  \label{r2}
\end{eqnarray}
where the projection of the bulk Weyl tensor to the surface
orthogonal to $n^{A}$ is given by \be E_{\mu \nu
}=\left.^{(5)}C_{RNS}^{M}\right.n_{M}n^{N}g_{\mu }^{R}g_{\nu
}^{S}\,. \label{weyl} \en Integrating out the five-dimensional
Einstein equation along the extra dimension from $y=-\epsilon $ to
$y=+\epsilon $, and taking the limit $\epsilon \rightarrow 0$,
leads to the Israel-Darmois junction conditions on the brane
$\lbrack q_{\mu \nu }]=0$ and $[K_{\mu \nu }]=-k_{5}^{2}(\tau
_{\mu \nu }-\frac{1}{3}g_{\mu \nu }\tau )$, where $[X]$ denotes
the jump of $X$ over the brane, $[X]=\lim_{y\rightarrow
+0}X-\lim_{y\rightarrow -0}X=X^{+}-X^{-}$. Furthermore, imposing
the Z$_{2}$-symmetry on the spacetime implies $K_{\mu \nu
}^{-}=-K_{\mu \nu }^{+}$, so that we can use the junction
condition equations to determine the extrinsic curvature on the
brane as \be K_{\mu \nu }=-\frac{1}{2}\,k_{5}^{2}\left(\tau _{\mu
\nu }-\frac{1}{3}\,g_{\mu \nu }\tau \right). \label{k4} \en
Finally, substituting this equation into Eq. (\ref{r2}), we obtain
the effective gravitational equations on the brane as
 \be G_{\mu \nu }=\frac{2k_{5}^{2}}{3}\left[T_{RS}^{(5)}\,g_{\mu
}^{R}g_{\nu }^{S}+g_{\mu \nu
}\left(T_{RS}^{(5)}n^{R}n^{S}-\frac{1}{4}\,T^{(5)}\right)\right]+k_{5}^{4}\pi
_{\mu \nu }-E_{\mu \nu }\,, \label{effectiveeq} \en where the
quadratic correction has the form \be \pi _{\mu \nu
}=-\frac{1}{4}\,\tau _{\mu \alpha }\tau _{\nu
}^{\alpha}+\frac{1}{12}\,\tau \tau _{\mu \nu }
+\frac{1}{8}\,g_{\mu \nu }\tau _{\alpha \beta }\tau ^{\alpha \beta
}-\frac{1}{24}\,g_{\mu \nu }\tau ^{2}\,, \label{pi3} \en and
$E_{\mu \nu}$ is given by Eq. (\ref{weyl}). The equations
(\ref{weyl}), (\ref{effectiveeq}) and (\ref{pi3}) are equivalent
to effective equations that were obtained in Ref.
\cite{Shiromizu:1999wj}. We assume that $L_{brane}$ includes
additional gravitational contributions of the form of induced
gravity on the brane. The only modification to take account in
this case is, in the spirit of Ref. \cite{Maeda:2003ar},
correction to the effective energy momentum tensor
$\tau_{\mu\,\nu}$. Recently, in Ref. \cite{Atazadeh:2007gs}, the
effective equations on the brane were found, where the induced
gravity was described by an arbitrary function $\mathcal{L}(R)$ of
the Ricci scalar. In this article the effective equations were
obtained by projection techniques. In the next section, we derive
the effective equations on the brane where the induced gravity
contains contributions of type $\mathcal{L}(R)$, using the
methodology developed in Ref. \cite{Maeda:2003ar}.

\section{\label{sec:level3} Effective equations on the brane; Modified
Gravity}

We consider induced gravity scenario that arises from higher-order
corrections on the scalar curvature over the brane. This scenario
corresponds to a version of the modified gravity on the DGP brane theory.
Our Lagrangian on the brane is given by
 \be
 L_{brane}=\frac{\mu
^{2}}{2}\,f(R)-\lambda +L_{m}\,,\label{lagrangian1}
 \en
 where $\mu$ is a mass scale which may correspond to the
4-D Planck mass, $\lambda$ is a tension of the brane and $L_{m}$
presents the Lagrangian of the matter fields onto the brane.
Higher-order correction to Einstein-Hilbert action typically arise
in low energy effective actions of string theory models. It is
well known that for $f(R)$ theories, if we make a conformal
transformation
$\overline{g}_{\mu\,\nu}=\Omega^{2}(x)\,g_{\mu\,\nu}$, are
equivalent to Einstein-Hilbert action non- minimal coupled to a
self-interacting scalar field which a potential obtained from the
conformal transformation. However, in our case situation is more
complicated because $f(R)$ term lives only on the brane, and the
conformal transformation should involves four dimensional sector
in order to change just the induce gravity. At this point, in
Ref.\cite{Nozari:2007qg} was studied  DGP inspired braneworld
scenario where a scalar field nonminimally coupled to the induced
Ricci curvature is present on the brane. Following the method
discussed in the previous section, we define the energy momentum
tensor as \be
 \tau _{\mu \nu }=-2\,\frac{\delta
L_{brane}}{\delta g^{\mu \nu }}+g_{\mu \nu
}L_{brane}\,.\label{tmunu}
\en
Using Eq. (\ref{lagrangian1}), the
energy momentum tensor reads
\begin{equation}
\tau _{\mu \,\nu }=-\Lambda (R)\,g_{\mu \,\nu }+T_{\mu \,\nu
}-\Sigma (R)\,G_{\mu \,\nu }+D_{\mu \,\nu },  \label{tmunu3}
\end{equation}%
where the functions $\Lambda(R)$, $\Sigma(R)$ and $D_{\mu \,\nu }$
are given by \be
\Lambda (R)=\frac{\mu ^{2}}{2}\left[R\,\frac{df}{dR}-f(R)+2\,\frac{%
\lambda }{\mu ^{2}}\right] \en and \be \Sigma (R)=\mu
^{2}\frac{df}{dR}\,, \en and
\be
D_{\mu \,\nu }=\mu ^{2}[D_{\mu
\,}D_{\nu }(\frac{df}{dR})-g_{\mu \,\nu }D^{\alpha }D_{\alpha
}(\frac{df}{dR})]
\en

 In this case, the quadratic corrections to
the Einstein equations on the brane are written as follow,
\begin{eqnarray}
\pi _{\mu \,\nu }&=&-\frac{1}{12}\Lambda ^{2}(R)\,g_{\mu \,\nu }+\frac{1}{6}%
\Lambda (R)\,T_{\mu \,\nu }-\frac{1}{6}\Lambda (R)\,\Sigma
(R)\,G_{\mu \,\nu }+\frac{1}{6}D_{\mu \,\nu }+\Sigma (R)^{2}\,\pi
_{\mu \,\nu }^{(G)}+\pi
_{\mu \,\nu }^{(T)}+\nonumber \\&&+\pi _{\mu \,\nu }^{(D)}+\Sigma (R)\,G^{\rho \,\sigma }\,%
\mathcal{K}_{\mu \,\nu \,\rho \,\sigma }^{(T)}+(\Sigma (R)G^{\rho
\,\sigma }-T^{\rho \sigma })\mathcal{K}_{\mu \,\nu \,\rho \,\sigma
}^{(D)}, \label{pimunu}
\end{eqnarray}
where we introduced the following definitions, \be
\pi_{\mu\,\nu}^{(G)}=-\frac{1}{4}\,G_{\mu\, \alpha }\,G_{\nu
}^{\alpha }+ \frac{1}{12}\,G\,G_{\mu \, \nu }+\frac{1}{8}\,g_{\mu
\,\nu }G_{\alpha\, \beta }G^{\alpha\, \beta
}-\frac{1}{24}\,g_{\mu\, \nu }G^{2}\,, \en and \be
\pi_{\mu\,\nu}^{(T)}=-\frac{1}{4}\,T_{\mu \,\alpha }\,T_{\nu
}^{\alpha }+\frac{1}{12}\,T\,T_{\mu\, \nu }+\frac{1}{8}\,g_{\mu\,
\nu }T_{\alpha\, \beta }T^{\alpha\, \beta }-\frac{1}{24}\,g_{\mu\,
\nu }T^{2}\,, \en
\begin{equation}
\pi _{\mu \,\nu }^{(D)}=-\frac{1}{4}D_{\mu \,\alpha }\,D_{\nu }^{\alpha }+%
\frac{1}{12}D\,D_{\mu \,\nu }+\frac{1}{8}g_{\mu \,\nu }D_{\alpha
\,\beta }D^{\alpha \,\beta }-\frac{1}{24}g_{\mu \,\nu }D^{2},
\end{equation}
 and
\begin{eqnarray}
\mathcal{K}_{\mu\,\nu\, \rho\, \sigma }^{(T)} &=&\frac{1}{4}\left(
g_{\rho \, \mu }T_{\sigma \, \nu }+g_{\sigma \, \nu }T_{\rho \,
\mu
}-g_{\mu \, \nu }T_{\rho \, \sigma }\right) + \nonumber \\
&&+\,\frac{1}{12}\left( T\left( g_{\mu \, \nu }g_{\rho \, \sigma
}-g_{\rho \, \mu }g_{\sigma º, \nu }\right) -T_{\mu \nu }g_{\rho
\, \sigma }\right),\label{k}
\end{eqnarray}
and
\begin{eqnarray}
\mathcal{K}_{\mu \,\nu \,\rho \,\sigma }^{(D)}
&=&\frac{1}{4}\left( g_{\rho \,\mu }D_{\sigma \,\nu }+g_{\sigma
\,\nu }D_{\rho \,\mu }-g_{\mu \,\nu
}D_{\rho \,\sigma }\right) +  \nonumber \\
&&+\frac{1}{12}\left( D\left( g_{\mu \,\nu }g_{\rho \,\sigma
}-g_{\rho \,\mu }g_{\sigma \nu }\right) -D_{\mu \nu }g_{\rho
\,\sigma }\right) ,
\end{eqnarray}
where $T$ is the trace of energy momentum tensor and $D$ is \be
D=g^{\mu \nu }D_{\mu \nu }=-3\mu ^{2}D^{\mu }D_{\mu
}(\frac{df}{dR}). \en

 Finally, replacing the quadratic corrections in the effective
Einstein equations (\ref{effectiveeq}), we obtain

\begin{eqnarray}
\left( 1+\frac{1}{6}k_{5}^{4}\Lambda (R)\,\Sigma (R)\,\right)
\,G_{\mu \,\nu
} &=&-\frac{k_{5}^{2}}{2}\Lambda ^{(5)}\,g_{\mu \,\nu }-\frac{1}{12}%
k_{5}^{2}\Lambda (R)^{2}\,g_{\mu \,\nu
}+\frac{1}{6}k_{5}^{4}\Lambda (R)T_{\mu \,\nu
}+\frac{1}{6}k_{5}^{4}D_{\mu \,\nu }+k_{5}^{4}\,\Sigma
(R)^{2}\,\pi _{\mu \,\nu }^{(G)}+k_{5}^{4}\,\pi _{\mu \,\nu
}^{(T)}+\nonumber\\&&+k_{5}^{4}\,\pi _{\mu \,\nu
}^{(D)}+k_{5}^{4}\Sigma (R)\,\,G^{\rho \,\sigma
}\,\mathcal{K}_{\mu \,\nu \,\rho \,\sigma }^{(T)}+k_{5}^{4}(\Sigma
(R)\,\,G^{\rho \,\sigma }\,-T^{\rho \,\sigma })\,\mathcal{K}_{\mu
\,\nu \,\rho \,\sigma }^{(D)}-E_{\mu \,\nu },\label{wa1}
\end{eqnarray}%
and the effective cosmological constant on the brane is given by
 \be
 \Lambda
^{\prime }=\frac{k_{5}^{2}}{2}\left(\Lambda
^{(5)}+\frac{1}{6}\,\Lambda(R) ^{2}k_{5}^{2}\right), \en note that
these equations are exactly the same effective equations as in the
reference \cite{Maeda:2003ar}, when $f(R)=R$.

\section{\label{sec:level3} Effective FRW Cosmological Equation on the
 Brane}

Now we are interested in computing the effective cosmological
equations on the brane. We discuss the Friedmann-Robertson-Walker
universes filled with a perfect fluid. First we consider the
five-dimensional metric described by the interval \be ds^{2}=d\chi
^{2}+g_{\mu \nu }dx^{\mu }dx^{\nu }, \label{metric4} \en and the
energy momentum tensor is given by $T_{AB}^{(5)}=\frac{-\Lambda
^{(5)}}{k_{5}^{2}}g_{AB}^{(5)}$. We decompose the metric as \be
 g_{AB}^{(5)}=
\left(\begin{array}{cc}
g_{\mu \nu } & 0 \\
0 & 1
\end{array}\right),
\en
so that its inverse has the form
\be
g^{(5)AB}=
\left(\begin{array}{cc}
g^{\mu \nu } & 0 \\
0 & 1
\end{array}\right),
\en and the normal vector to the brane is $n^{R}=(0,0,0,0,1)$.
Therefore, it is not hard to show that the effective equations
(\ref{effectiveeq}) reduce to \be
\frac{2}{3}\,k_{5}^{2}\left[T_{RS}^{(5)}\,g_{\mu }^{R}g_{\nu
}^{S}+g_{\mu \nu
}\left(T_{RS}^{(5)}\,n^{R}n^{S}-\frac{1}{4}\,T^{(5)}\right)\right]=-%
\frac{1}{2}\,\Lambda ^{(5)}g_{\mu \nu }\,, \en and therefore the
basic equations read as follows,
\begin{eqnarray}
 G_{0}^{0}&=&-\frac{1}{2}\,\Lambda ^{(5)}+k_{5}^{4}\pi
_{0}^{0}-E_{0}^{0}\,,\nonumber \\
G_{j}^{i}&=&-\frac{1}{2}\,\Lambda ^{(5)}\delta
_{j}^{i}+k_{5}^{4}\pi _{j}^{i}-E_{j}^{i}\,. \label{basiceq}
\end{eqnarray}
On the other hand, from the geometry of Friedmann-Robertson-Walker
universes, it is well-known that the components of the Einstein
tensor are given by the following expressions,
\begin{eqnarray}
G_{0}^{0}&=&-3\left(H^{2}+\frac{k}{a^{2}}\right)\,, \nonumber \\
G_{j}^{i}&=&-\left(2H^{\prime }+3H^{2}+\frac{k}{a^{2}}\right)\delta _{j}^{i}\,,
\label{basiceq2}
\end{eqnarray}
where $H=\frac{\dot{a}}{a}$ represents the Hubble constant, $a(t)$
is the scale factor and $k$ describes the  hypersurfaces of
homogeneity that could be represented as a three-sphere, a
three-plane or a three-hyperboloid, with values $k=1, 0,-1$,
respectively. Since the space-time is isotropic and homogeneous,
it is possible to show that $D^{\nu}\,\pi_{\mu\,\nu}=0$ and
$D^{\nu}\,E_{\mu\,\nu}=0$ are satisfied \cite{Shiromizu:1999wj},
and the relevant components of $\pi_{\mu\,\nu}$ are given by
\begin{eqnarray}
\pi _{0}^{0}&=&-\frac{1}{12}\,(\tau _{0}^{0})^{2} \nonumber \\
\pi _{j}^{i}&=&\frac{1}{12}\,\tau _{0}^{0}(\tau _{0}^{0}-2\tau
_{1}^{1})\delta _{j}^{i}\,.\label{basicpi}
\end{eqnarray}
The energy momentum tensor then becomes
\begin{equation}
\tau _{\nu }^{\mu }=-\mu ^{2}\frac{df}{dR}G_{\nu }^{\mu }-\frac{\mu ^{2}}{2}%
\left[ R\frac{df}{dR}-f(R)+2\frac{\lambda }{\mu ^{2}}\right]
\delta _{\nu
}^{\mu }+T_{\nu }^{\mu }+\mu ^{2}[D_{\,}^{\mu }D_{\nu }(\frac{df}{dR}%
)-\delta _{\,\nu }^{\mu }D^{\alpha }D_{\alpha }(\frac{df}{dR})],
\label{tmunu1}
\end{equation}
where $T_{\nu }^{\mu }=diag(-\rho ,P,P,P)$. Combining 00 component
of Eqs. (\ref{basiceq}), (\ref{basiceq2}) and (\ref{tmunu1}),
solving the equation obtained for $H^2+k/a^2$, we obtain the
effective basic cosmological equations on the brane,
\begin{eqnarray}
H^{2}+\frac{k}{a^{2}}&=&\frac{2}{\mu^{4}k_{5}^{4}(\frac{df}{dR})^{2}}
+\frac{1}{3\mu ^{2}(\frac{df}{dR})}\left[\frac{\mu
 ^{2}}{2}\left(R\frac{df}{dR}
-f(R)-6H\dot{R} \frac{d^2f(R)}{dR^2}\right)+\lambda +\rho \right] \nonumber \\
&&\pm \frac{2}{k_{5}^{2}\mu ^{2}(\frac{df}{dR})}\sqrt{
\frac{1}{k_{5}^{4}\mu ^{4}(\frac{df}{dR})^{2}} +\frac{\frac{\mu
^{2}}{2}\left(R\frac{df}{dR}-f(R)-6H\dot{R}
\frac{d^2f(R)}{dR^2}\right)+\lambda +\rho }{3\mu
 ^{2}(\frac{df}{dR})}-\frac{\Lambda^{(5)}}{6}-\frac{\varepsilon
_{0}}{3a^{4}}}\,, \label{hgigante}
\end{eqnarray}
where $\varepsilon _{0}$ is an integration constant. The above
equation represents the effective Friedmann equation for the DGP
theory with curvature corrections on the brane. Note that correct
limit is obtained when $f(R)=R$. In order to study the different
limits of Eq. (\ref{hgigante}), we combined 00 component of Eqs.
(\ref{basiceq}), (\ref{basiceq2}) and (\ref{tmunu1}), then taking
root-sqare of the obtained equation, we arrive to more manipulable
form \cite{Brown:2007dd}
%%%%%%%%%%%%%%%%%%%%%%%%%%%%%%%%%%%%%%%%%%%%%%%%%%%%%%%%%%%%%%%%%%%%%%%%
\begin{equation}
r\left[ \left( H^{2}+\frac{k}{a^{2}}\right) \frac{df}{dR}\right] -\frac{%
k_{5}^{2}}{6}\left( \rho +\lambda +\frac{\mu ^{2}}{2}\left(R\frac{df}{dR}%
-f(R)-6H\dot{R} \frac{d^2f(R)}{dR^2}\right)\right) =\pm
\sqrt{H^{2}+\frac{k}{a^{2}}-\frac{1}{3}\frac{\varepsilon
_{0}}{a^{4}}-\frac{1}{6}\Lambda ^{(5)}},\label{hutil}
\end{equation}
where $r$ is the cross-over scale and is defined as
$r=\frac{1}{2}k_{5}^{2}\mu
^{2}=\frac{1}{2}\frac{k_{5}^{2}}{k_{4}^{2}}$, and two different
solutions ($\pm )$ correspond to two different embedding of the
brane within the bulk..

If we choose a Minkowskian bulk ($\Lambda ^{(5)}=0$, $\varepsilon
_{0}=0$ ) with  $\lambda =0$ on a flat brane ($k = 0$) we get
\begin{equation}
rH^{2}\frac{df}{dR}-\frac{k_{5}^{2}}{6}\left( \rho +\frac{\mu ^{2}}{2}(R%
\frac{df}{dR}-f(R)-6H\dot{R}\frac{d^{2}f}{dR^{2}})\right) =\pm H
\label{ecuacion 100}
\end{equation}
where for $f(R)=R$ (DGP standard model) we obtain the right
limits, at low energies $\rho\longrightarrow 0$ appear the brach
DGP(+) characterized by $H\longrightarrow \frac{1}{r}$  and the
branch DGP(-) here $H\longrightarrow 0$. DGP(+) branch is
important from cosmological point of view, because this branch has
a self-accelerated phase. Note that this self-acceleration in this
limit can be seen in term of scale factor as an exponential
expansion $a(t)=a_{0}e^{\frac{t}{r}}$. In order to explore
cosmological consequence we are considering $f(R)=R^{m}$, and also
we focus at low energies limit ($\rho \longrightarrow 0$) in a
Minkowskian bulk $\Lambda ^{(5)}=0$, $\varepsilon _{0}=0$, and the
case of flat brane with $\lambda =0$. We are asking if our theory
allows a solution of the form $a(t)=a_{0}e^{\alpha (t-t_{0})}$,
where $\alpha$ is a constant parameter. Then if we replace our
choice of $f(R)$ and the Ansatz of the scale factor in
Eq.(\ref{ecuacion 100})\footnote{where we used that
$R=12H^{2}+6\dot{H}=12\alpha ^{2}$} we obtain,
\begin{equation}
r12^{m-1}\alpha ^{2m}(2-m)=\pm \alpha,  \label{ecuaicon 104}
\end{equation}
this equation allows to investigate behaviors of the solution in
this modified DGP brane. First we consider the case $m=1$ (DGP
standard brane), we obtain $H=\alpha =\frac{1}{r}$ positive branch
and $H=\alpha =0$ for the negative branch(-), therefore we recover
standard behaviors of DGP brane. In the case $m=2$ we obtain
$H=\alpha =0$ in both branches static universe (like Einstein's
universe). For $m>2$ we obtain two solutions for the Hubble
parameter $H=\alpha =0$  the static branch and \be H=\alpha =\left( \frac{1}{(m-2)r12^{m-1}}\right) ^{\frac{1}{(2m-1)}%
}\en for the accelerated branch. Therefore  for $f(R)=R^m$ it is
possible to obtain a model with self-acceleration for $m>0$
($m\neq2$). The self-acceleration for the case with $m=2$, it is
included if we choose $f(R)=R+\beta R^m$. In sum we obtain a
self-accelerated cosmology at the low energy limit.

On the other hand for  the high energies limit
$H>>\frac{1}{r\frac{df}{dR}}$ of Eq. (\ref{ecuacion 100})for  a
Minkowskian bulk $\Lambda ^{(5)}=0$, $\varepsilon _{0}=0$, and the
case of flat brane with $\lambda =0$,  we obtain that both
branches have the same limit \be
H>>\frac{1}{r\frac{df}{dR}},\Longrightarrow H^{2}=\frac{1}{3\mu ^{2}\frac{df%
}{dR}}\left( \rho +\lambda +\frac{\mu ^{2}}{2}(R\frac{df}{dR}-f(R)-6H\dot{R}%
\frac{d^{2}f}{dR^{2}})\right), \label{hel}\en where we recover the
standard Friedmann equation if we take $f(R)=R$. We noticed that
Eq. (\ref{hel}) is equal to one that is obtained from $4D$
modified gravity \cite{Fay:2007uy}. Therefore the 4D regime is
driven by a modified gravitational theory.

Finally as $R$ contains second derivatives of the scale factor
$a(t)$ in Eq.(\ref{hgigante}), means that there are more degrees
of freedom in the space of solutions than for Einstein gravity
with a cosmological constant or the standard DGP model and it is
needed to confront this theory with observational data. At this
point, one interesting approach is to make a conformal
transformation in to the four dimensional part of the theory.
Where we can see explicitly the new degree of freedom thought the
scalar field that appear in the transformed theory. This theory
behaviors like a DGP model with a scalar field ($\phi$) (where are
disguise all the effect of the modified gravity in DGP theory)
with a potential $V(\phi)$ evolving in FRW DGP brane where the
matter is endowed with a non-minimal coupling set by $\phi$.
Certainly, Eq. (\ref{hgigante}) encodes all cosmological
information that  we hope to discuss in the near future. In sum
using the method developed in Ref. \cite{Maeda:2003ar}, we obtain
the effective gravitational equations on the DGP brane world where
the induced gravity was described by  a general $f(R)$ term.
Applying the formulas to cosmology --a universe filled with a
perfect fluid--, we derive the generalized Friedmann equation on
the brane. This modified Friedmann equation give us a new
possibility for studying cosmological models through the inclusion
of curvature corrections in the induced gravity term on the brane,
and really open a new kind of model still unexplored.
 In this sense it is interesting to consider both: the DGP model and modified gravity
as an unified scenario that we hope could be able to describe the
majority of cosmological puzzles: the late acceleration,
unification of early acceleration (inflation) with dark energy,
also unification of the dark matter and dark energy and maybe a
natural scheme that provide a mechanism in order to describes the
transition from decceleration to acceleration in the universe
evolution. . We note that our gravitational effective equations
(\ref{effectiveeq}), as well as the Friedmann effective equation,
have the correct limit when $f(R)=R$, and can be considered as a
genuine generalization of DGP effective equations on modified DGP
brane world.  Whose high and low energies limits show a
self-accelerated phase (interesting from the cosmological point of
view) in the $5D$ regime and a modified gravitational theory for
the $4D$ regime, respectively. Investigations of further
cosmological implications, including the late acceleration of the
universe, are under our current considerations.

\begin{acknowledgments}
We are grateful to A.~N.~Aliev,O. Mi\v{s}kovi\'{c} and
S.~D.~Odintsov for enlightening comments.  J.S. is supported by
the COMISION NACIONAL DE CIENCIAS Y TECNOLOGIA through FONDECYT
Grant N$^{0}$. 11060515, and also was partially supported by PUCV
Grant N$^0$. 123.789/2007. Y.V was partially supported by
MINISTERIO DE EDUCACION through MECESUP Grants FSM 0204, by
Direcci\'on de Estudios Avanzados PUCV and by CONICYT Scholarship
2008.
\end{acknowledgments}

\end{document}